\documentclass{svjour3}
\usepackage{graphicx,latexsym}
\usepackage[noadjust]{cite}
\usepackage{amsmath} % assumes amsmath package installed
\usepackage{amssymb}  % assumes amsmath package installed
\usepackage{url}

\smartqed

\def\proof{{\em Proof: }}

\def\C{{\mathbb C}}

\begin{document}
\title{On graphs whose Laplacian matrix's multipartite separability is invariant under graph isomorphism}

\author{Chai Wah Wu}
\institute{C. W. Wu \at IBM T. J. Watson Research Center\\
P. O. Box 704, Yorktown Heights, NY 10598, USA\\
\email{cwwu@us.ibm.com,chaiwahwu@member.ams.org}}

\date{}

\maketitle

\begin{abstract}
Normalized Laplacian matrices of graphs have recently been studied in the context of quantum mechanics as density matrices of quantum systems.  Of particular interest is the relationship between quantum physical properties of the density matrix and the graph theoretical properties of the underlying graph.  One important aspect of density matrices is their entanglement properties, which are responsible for many nonintuitive physical phenomena.  The entanglement property of normalized Laplacian matrices is in general not invariant under graph isomorphism.  In recent papers, graphs were identified whose entanglement and separability properties are invariant under isomorphism.  The purpose of this note is to characterize the set of graphs whose separability is invariant under graph isomorphism.  In particular, we show that this set consists of $K_{2,2}$, $\overline{K_{2,2}}$ and all complete graphs.

\end{abstract}

\section{Introduction}
The objects of study in this paper are density matrices of quantum mechanics.  Density matrices are used to describe the state of a quantum system and are fundamental mathematical constructs in quantum mechanics.  They play a key role in the design and analysis of quantum computing and information systems \cite{nielsen-quantum-2002}.

\begin{definition}
A complex matrix $A$ is a {\em density matrix} if it is Hermitian, positive semidefinite and has unit trace.
\end{definition}

\begin{definition}
A complex matrix $A$ is row diagonally dominant if $A_{ii} \geq \sum_{j\neq i} |A_{ij}|$ for all $i$.
\end{definition}

By Gershgorin's circle criterion, all the eigenvalues of a row diagonally dominant matrix has nonnegative real parts.  Thus a nonzero Hermitian row diagonally dominant matrix is positive semidefinite and has a strictly positive trace, and such a matrix normalized\footnote{We define the normalization of a matrix $A$ with nonzero trace as $\frac{1}{\mbox{tr}(A)}A$.} is a density matrix.

A key property of a density matrix is its separability.  The property of nonseparability is crucial in generating the myriad of counterintuitive phenomena in quantum mechanics and is indispensable in the construction of quantum information processing systems.

\begin{definition}
A density matrix $A$ is {\em separable} in $\C^{p_1}\times \C^{p_2} \times \cdots \times \C^{p_m}$ ($p_i\geq 2$)
if it can be written as $A = \sum_i c_i A_i^1 \otimes \cdots \otimes A_i^m$ where
$c_i\geq 0$, $\sum_i c_1 = 1$ and $A_i^j$ are density matrices in $\C^{p_j\times p_j}$.
A density matrix is {\em entangled} if it is not separable.
\end{definition}

\section{Laplacian matrices as density matrices}
The Laplacian matrix of a graph is defined as $L = D-A$ , where $D$ is the diagonal matrix of the vertex degrees and $A$ is the adjacency matrix.  The matrix $L$ is symmetric and row diagonally dominant, and therefore for a nonempty\footnote{A graph is {\em empty} if it has no edges.  In this case the Laplacian matrix is the zero matrix and has zero trace.} graph the matrix $\frac{1}{\mbox{tr}(L)}L$ is a density matrix.
In Ref. \cite{braunstein:laplacian:2006}, such normalized Laplacian matrices are studied as density matrices and quantum mechanical properties such as entanglement of various types of graph Laplacian matrices are studied.
This approach was further investigated in \cite{wu:separable:2006} where it was shown that the Peres-Horodecki necessary condition for separability is equivalent to a condition on the partial transpose graph, and that this condition is also sufficient for separability of block tridiagonal Laplacian matrices and Laplacian matrices in $\C^2\times \C^q$.    In Ref. \cite{braunstein:laplacian_graph:2006} several classes of graphs were identified whose separability are easily determined.   In Ref. \cite{wang:tripartite:2007} the tripartite separability of normalized Laplacian matrices is studied.  Clearly separability and entanglement are only nontrivial if the order of the density matrix is composite.  Therefore in this note we only consider nonempty graphs on $n$ vertices where $n$ is a composite integer.

As noted in Ref. \cite{braunstein:laplacian:2006}, the separability of a normalized Laplacian matrix of a graph is not invariant under graph isomorphism; it depends on the labeling of the vertices.  In the sequel, unless otherwise noted, we will assume a specific Laplacian matrix (and thus a specific vertex labeling) when we discuss separability of Laplacian matrices of graphs.
To determine separability of normalized Laplacian matrices, it is more convenient for a graph of
$n=p_1p_2\cdots p_m$ vertices to consider the vertices as $m$-tuples in $V_1\times V_2\times\cdots\times V_m$, $|V_i| = p_i$.
In particular, we define a vertex labeling as:

\begin{definition}
For $n= p_1p_2\cdots p_m$, a {\em vertex labeling} is a bijection between
$\{1,\dots , n\}$ and $\{1,\dots , p_1\}\times \{1,\dots , p_2\} \times \cdots \times
\{1,\dots ,p_m\}$.
\end{definition}

\begin{definition}
Given a graph $\cal G$ with vertices  $V\times W$, the partial transpose graph ${\cal G}^{pT}$ is a graph with vertices
$V\times W$ and edges defined by:

$\{(u,v),(w,y)\}$ is an edge of $\cal G$ if and only if $\{(u,y),(w,v)\}$ is an edge of ${\cal G}^{pT}$.
\end{definition}
Note that the partial transpose graph depends on the specific labeling of the vertices.  The partial transpose graph is useful in determining separability of the Laplacian matrix of a graph with the same vertex labeling.
In \cite{braunstein:laplacian_graph:2006,wu:separable:2006}
the following necessary condition for separability is shown:

\begin{theorem} \label{thm:separable}
If the normalized Laplacian matrix of $\cal G$ is separable then each vertex of $\cal G$ has the same degree as the same vertex of ${\cal G}^{pT}$. \label{thm:degree}
\end{theorem}

In  Ref. \cite{wu:multipartite:2009} several classes of graphs were identified where the normalized Laplacian matrices' separability or entanglement is invariant under graph isomorphism.  In particular, it was shown that for all noncomplete graphs of $2m$ vertices $m\geq 3$, there is a vertex labeling that render the normalized Laplacian matrix entangled.  Computer experiments performed in Ref. \cite{wu:multipartite:2009} indicate that this is true for all noncomplete graphs with $4 < n \leq 9$ vertices where $n$ is composite.  The purpose of this note is to show that this is indeed the case for all $n>4$ and this allows us to characterize completely the set of graphs whose separability is invariant under graph isomorphism.

\section{Graphs whose normalized Laplacian matrix is multipartite separable for all vertex labelings} \label{sec:separable}
We will use the following generalization of the Pigeonhole Principle:
\begin{theorem}[Generalized Pigeonhole Principle]
\label{thm:gpp}
If $rn+s$ or more objects are placed in $n$ boxes, where $0\leq s<n$,
then for each
$1\leq m\leq n$, there exists $m$ boxes with a total of at least $rm+\min(s,m)$ objects.
\end{theorem}
\proof We prove this by induction on $m$.
First consider the case $m=1$.  This case is well known, but we include the proof here for completeness. For $s=0$, if all boxes has $r-1$ or less objects, then
there is a total of at most $(r-1)n$ objects, a contradiction.
For $s\geq 1$, if all boxes have $r$ or less objects then there is a total of at most $rn$ objects which is strictly less than $rn+s$, a contradiction. This case of $m=s=1$ is also known as the Extended  Pigeonhole Principle \cite{bogart:combinatorics:2000}.

Assume the theorem is true for $m=k$ for some $1\leq k < n$. 
Then there exists $k$ boxes with a total of $rk+u$ objects where $u\geq \min(k,s)$.  
The number of objects in the remaining $n-k$ boxes is $rn+s-rk-u = r(n-k) - (u-s)$.

{\bf Case 1}: $u = s$. By the $m=1$ case, there exists a box in the remaining boxes with $r$ objects.  This means that we have $k+1$ boxes with at least
$rk+u+r = r(k+1) + s \geq r(k+1)+ \min(s,k+1)$ objects.

{\bf Case 2}: $u > s$.
Let $r(n-k) - (u-s) = l(n-k) + t$, $r>l\geq 0$, $0\leq t \leq n-k-1$.
Then by the $m=1$ case there exists a box with at least $l+\min(t,1)$ objects.
Thus this box with the previously identified $k$ boxes form $k+1$ boxes with at least
$w = rk+u+l+\min(t,1) \geq r(k+1) - (r-l) + u$ objects.  Since $u = (r-l)(n-k)+s-t\geq (r-l-1)(n-k)+s+1$, this means that $w \geq r(k+1) + (r-l-1)(n-k-1) + s \geq r(k+1)
+ \min(s,k+1)$.

{\bf Case 3}: $s > u$.  In this case $s > k$, $u\geq k$ and
$rn+s-rk-u> r(n-k)$.   This implies that there is a box with at least $r+1$ objects in the remaining $n-k$ boxes. Adding this to the $k$ boxes results in $k+1$ boxes with at least $rk+u+r+1 \geq r(k+1) + (k+1) = r(k+1) + \min(s,k+1)$ objects.

In each case the theorem is true for $m=k+1$ and the proof is complete.\qed

Our main result is the following:
\begin{theorem}
The only graphs whose normalized Laplacian matrix is multipartite separable for all vertex labelings are $K_{2,2}$, $\overline{K_{2,2}}$ and the complete graphs $K_n$.
\end{theorem}
\proof In Ref. \cite{wu:multipartite:2009} it was shown that the graphs described in the theorem are multipartite separable for all vertex labelings.  Thus it suffices to show that all other graphs are multipartite entangled for some vertex labeling.  To this end, we find a vertex labeling which violates the vertex degree condition in 
Theorem \ref{thm:separable}.  

The case of $\C^{2}\times \C^{2}$ can be checked by explicit enumeration since the vertex degree condition in Theorem \ref{thm:separable} is both sufficient and necessary \cite{wu:separable:2006} and it shows that only $K_4$, $K_{2,2}$ and $\overline{K_{2,2}}$ has separable Laplacian matrices for all vertex labelings.  
Next we consider
$\C^{p_1}\times \C^{p_2} \times \cdots \times \C^{p_m}$ where $p_i\geq 3$ for some $i$.
Since entanglement in $\C^{p_1}\times \C^{p_2p_3\cdots p_m}$ implies
entanglement in $\C^{p_1}\times \C^{p_2} \times \cdots \times \C^{p_m}$, we need only to consider the case $\C^{p}\times \C^{q}$ where $p\geq 3$. Consider a graph that is not complete.
Let $d$ be the minimal vertex degree of the graph.   The case $d< q$ has already been proven in 
Ref. \cite{wu:multipartite:2009}, so we assume $d \geq q$.  
Define $k = q(p-1) = n-q$ and $t = q(d-q+1)$.  Note that $t > 0$ since $d\geq q$.
Let us denote $V$ as the set of vertices.  Pick a vertex $x$ with degree $d$.  Our goal is to find two disjoint subsets of vertices $U$, $W$ of size $q$ and $p-1$ respectively, such that
$x\in U$ and $e(\{x\},U) + e(U,W) > d$, where we use $e(A,B)$ to denote the number of edges 
between vertices in $A$ and $B$.  Let $t = rk+s$ where $r\geq 0$ and
$0\leq s<k$.  We consider $3$ cases.

{\bf Case 1}: $0<s<p-1$. Let $U$ be the set of vertices consisting of $x$ and $q-1$ vertices adjacent to $x$.  Since each vertex in $U$ has degree at least $d$, the number of edges between $U$ and $V\backslash U$ is at least $q(d-q+1) = t$.
Since $|V\backslash U| = k$, by Theorem \ref{thm:gpp} there exists $p-1$ vertices $W$ in $V\backslash U$ with at least
$r(p-1)+s$ edges to $U$.  The number of edges from $x$ to vertices in $U$ is $q-1$.
Since $s > 0$, $qr(p-1) +sq = kr + sq = t+s(q-1) > t$.
This implies $r(p-1) + s > t/q = d-q+1$.

{\bf Case 2}: $p-1\leq s<k$.  We pick $U$ as in Case 1.  By Theorem \ref{thm:gpp} there exists at $p-1$ vertices $W$ in $V\backslash U$ with at least
$(r+1)(p-1)$ edges to $U$.  Then $q(p-1)(r+1) = k(r+1) = t-s+k > t$ since $k>s$.
This means that $(p-1)(r+1) > t/q = d-q+1$.

{\bf Case 3}: $s = 0$. This implies that $r>0$ and $rk = t = q(d-q+1)$. Since  the graph is not complete $d< n-1$ and there is a vertex $y$ not adjacent to $x$.  Let $U$ be $x$, $y$ and $q-2$ vertices adjacent to $x$.
The number of edges from $x$ to $U$ is $q-2$ whereas 
$e(U,V\backslash U) \geq (q-2)(d-q+1) + 2(d-q+2) = q(d-q+1) + 2 = rk+2$.
Since $p\geq 3$, by Theorem \ref{thm:gpp} we can pick $p-1$ vertices $W$ with at least $r(p-1)+2$ edges to $U$.
Note that $qr(p-1)+q = kr + q > kr$.  This means that
$r(p-1) + 2 > kr/q + 1 = d-q+2$.

So in each case we have
$e(\{x\},U) + e(U,W) > d$.
For the vertex labeling we assign $(v_1,w_i)$ to the vertices in $U$ with vertex $x$ equal to $(v_1,w_1)$ and assign
$(v_j,w_1)$ to vertices in $W$, with $j\geq 2$.  Each edge between $x$ and a vertex in $U$ will remain an edge in the partial transpose graph.
Each edge between a vertex in $U$ and a vertex in $W$ will be an edge in the partial transpose graph as well.  This means that $x$ has degree $e(\{x\},U) + e(U,W) > d$
in the partial transpose graph and thus by Theorem \ref{thm:separable} the
normalized Laplacian matrix is entangled for this vertex labeling.\qed

\section{Conclusions}
We continue the study of normalized Laplacian matrices of graphs as density matrices and analyze their entanglement properties.  We can classify the set of graphs into 3 categories.  

{\bf Class S}: graphs whose normalized Laplacian matrix is separable under all vertex labelings.
In this note we show that this class of graphs consists of the complete graphs, $K_{2,2}$ and $\overline{K_{2,2}}$.  

{\bf Class SE}: graphs for which there is a vertex labeling such that the normalized Laplacian matrix is separable and there is a vertex labeling such that the normalized Laplacian matrix is entangled.  For separability in $\C^{2}\times \C^{2}$ and $\C^{2}\times \C^{3}$ this class of graphs were listed in Ref. \cite{wu:multipartite:2009}.
Furthermore, it was shown that for separability in $\C^{p}\times \C^{q}$, $pq > 4$,
nonempty complete bipartite graphs $K_{r,n-r}$ and their complements $\overline{K_{r,n-r}}$ where $r\equiv 0 \mod p$
belong to class SE.

{\bf Class E}: graphs whose normalized Laplacian matrix is entangled under all vertex labelings.
For separability in $\C^{p_1}\times \C^{p_2} \times \cdots \times \C^{p_m}$, it was shown in Ref. \cite{wu:multipartite:2009} that $K_{r,n-r}$ and $\overline{K_{r,n-r}}$ where $1\leq r< \frac{n}{p_1}$ and $r\not\equiv 0 \mod p_1$ belong to class E.

In general, characterizing the graphs in class SE and class E is still an open problem.

\bibliographystyle{IEEEtran}
\bibliography{quantum,graph_theory,algebraic_graph,math}
\end{document}